\documentstyle[aas2pp4]{article} 
\begin{document}
\title {OXYGEN ABUNDANCES IN METAL-POOR STARS
($-$2.2 $<$ [Fe/H] $<$ $-$1.2) FROM INFRARED OH LINES}

\newcommand{\teff}{T$_{\rm eff}$ }
\newbox\grsign \setbox\grsign=\hbox{$>$} \newdimen\grdimen \grdimen=\ht\grsign
\newbox\simlessbox \newbox\simgreatbox
\setbox\simgreatbox=\hbox{\raise.5ex\hbox{$>$}\llap
     {\lower.5ex\hbox{$\sim$}}}\ht1=\grdimen\dp1=0pt
\setbox\simlessbox=\hbox{\raise.5ex\hbox{$<$}\llap
     {\lower.5ex\hbox{$\sim$}}}\ht2=\grdimen\dp2=0pt
\def\simgreat{\mathrel{\copy\simgreatbox}}
\def\simless{\mathrel{\copy\simlessbox}}
\newbox\simppropto
\setbox\simppropto=\hbox{\raise.5ex\hbox{$\sim$}\llap
     {\lower.5ex\hbox{$\propto$}}}\ht2=\grdimen\dp2=0pt
\def\simpropto{\mathrel{\copy\simppropto}}

\author{\bf Jorge Mel\'endez
{\footnote{Visiting Astronomer, Kitt Peak National Observatory,
National Optical Astronomy Observatories, which is operated by the
Association of Universities for Research in Astronomy, Inc. (AURA) 
under cooperative agreement with the National Science Foundation}} 
and Beatriz Barbuy}
\affil{Universidade de S\~ao Paulo, IAG, Dept. de Astronomia,
 CP 3386, S\~ao Paulo 01060-970, Brazil \\ E-mail:
jorge@iagusp.usp.br, barbuy@orion.iagusp.usp.br}

\and

\author{\bf Fran\c cois Spite}
\affil{Observatoire de Paris-Meudon, DASGAL, F-92195 Meudon Cedex, France \\
E-mail: Francois.Spite@obspm.fr}

\slugcomment{Submitted to the The Astrophysical Journal}
\slugcomment{Send proofs to:  J. Melendez}

\begin{abstract}

Infrared OH lines at 1.55 - 1.56 $\mu$m in the H-band were 
obtained with the Phoenix high-resolution spectrograph
at the 2.1m telescope of the Kitt Peak National Observatory
for a sample of 14 metal-poor stars.

Detailed analyses of the sample stars have been carried out,
deriving stellar parameters based on two methods:
(a) spectroscopic parameters; (b) IRFM effective temperatures,
trigonometric gravities and metallicities from \ion{Fe}{2} lines.
The \ion{Fe}{1} lines present in the H-band region observed showed 
to be well fitted by the stellar parameters 
within $\Delta$[Fe/H] $\leq$ 0.15 dex.

The oxygen abundances were derived from fits of
spectrum synthesis calculations to the infrared OH lines.
CO lines in the H- and K-bands were obtained for a subsample
in order to determine their carbon abundances.

Adopting the spectroscopic parameters a mean 
oxygen-to-iron ratio of [O/Fe] $\approx$ +0.52 is obtained,
whereas using the IRFM temperatures, Hipparcos gravities and [\ion{Fe}{2}/H],
[O/Fe] $\approx$ +0.25 is found. A mean of the two methods gives
a final value of [O/Fe] $\approx$  +0.4 
for the metallicity range $-$2.2 $<$ [Fe/H] $<$ $-$1.2 
of the sample metal-poor stars.

\end{abstract}

\section{Introduction}

Oxygen abundances in metal-poor stars are a key result for constraining 
several important issues such as the age of globular clusters 
(VandenBerg 1985, VandenBerg et al. 2000), models of cosmic ray spallation
(Walker et al. 1993; Fields \& Olive 1999;
Kneller, Steigman \& Walker 2001), chemical evolution
models in the early phases of the Galaxy evolution
(Chiappini et al. 1999, hereafter CMBN99; 
Pagel \& Tautvai\v{s}ien\.{e} 1995, hereafter PT95) 
and nucleosynthesis
ejecta of massive stars (Thielemann, Nomoto \& Hashimoto 1996, hereafter
TNH96;
Woosley \& Weaver 1995, hereafter WW95).

Oxygen is a major tracer of chemical evolution, since it is
a bona fide primary element ejected 
by massive stars. The oxygen-to-iron ratio is a measure of SNII/SNI 
rates along time, and can be used to characterize fast chemical 
enrichment ([O/Fe] $>$ 0) such as in the halo of our Galaxy.

There is agreement in the literature concerning an overabundance of 
oxygen relative to iron in metal-poor stars.
This result was first established by Conti et al. (1967) and
Sneden et al. (1979). However there is no 
agreement on the overabundance value itself. The reasons
for the discrepancies come from the fact that
 only four sets of lines can be used to derive oxygen abundances
in metal-poor stars: 
(i) the forbidden [\ion{O}{1}]$\lambda$6300.31, $\lambda$6363.79 {\rm \AA}
 lines measurable in giants of [Fe/H] $>$  $-$3.0;
(ii) the permitted \ion{O}{1} $\lambda$ 7771.96, 7774.18 and 7775.40 
{\rm \AA} lines  measurable in dwarfs and subgiants; 
(iii) the ultraviolet (UV) OH lines (A$^2 \Sigma - $X$^2 \Pi$ electronic transition);
(iv) the infrared (IR) OH lines (X$^2 \Pi$ vibration-rotation transition).

Therefore there are only a few lines available, and in most cases
the different lines are not present in the same stars. 
\ion{O}{1} permitted lines give systematically higher values than
the [\ion{O}{1}] forbidden lines. The problems with the different lines
can be summarized as follows.

(i) The reliability of the forbidden lines [\ion{O}{1}] present in giants
 stems from the fact that the transition involves a metastable level, 
collisionally controlled, therefore not directly subject to non-LTE effects 
(see Kiselman 2001a,b; Lambert 2001).
Values of [O/Fe] $\approx$ +0.4 to +0.6 
 down to [Fe/H] $\approx$ $-$3.0 are obtained
(e.g. Gratton \& Ortolani 1986; Barbuy 1988; Sneden et al. 1991;
Spiesman \& Wallerstein 1991; Kraft et al. 1992; Shetrone 1996;
 Carretta et al. 2000; Gratton et al. 2000; Westin et al. 2000;
Asplund 2001a,b; Cayrel et al. 2001a; Nissen et al. 2001;
Sneden \& Primas 2001a,b); 
below  [Fe/H] $<$ $-$3.0 the oxygen forbidden lines are too faint.

(ii) the permitted lines of the \ion{O}{1} triplet present in dwarfs, 
give [O/Fe] $\approx$ +1.0 at [Fe/H] $\approx$ $-$3.0 
(e.g. Abia \& Rebolo 1989;
Cavallo, Pilachowski \& Rebolo 1997; Mishenina et al. 2000). 
However the permitted lines of high excitation potential seem
to be subject to overenhancements, partly due to non-LTE
(Kiselman 2001a,b) and partly 
to uncertainties in the deep layers of model atmospheres,
as is also the case of \ion{C}{1} lines of high excitation potential
(Tomkin et al. 1992). An  evidence of this 'artificial' overenhancement 
was demonstrated through the study of both forbidden and permitted lines 
in some samples of stars, and they give different results for a given star 
(Spite \& Spite 1991; Fulbright \& Kraft 1999).

King (1993) claimed that by adopting a hotter effective 
temperature T$_{\rm eff}$ scale and depending on  the models,
it is possible to obtain
 [O/Fe] $\approx$ +0.5 for halo dwarfs derived from
the permitted lines,
comparable to those obtained from the forbidden lines for giants.
Carretta et al. (2000) obtained the same results, also adopting
a higher T$_{\rm eff}$ scale and non-LTE corrections applied
to the permitted lines. King (2000) applied NLTE corrections
on the stellar parameters log $g$ and [Fe/H], and obtained
[O/Fe] $\approx$ 0.6 at [Fe/H] $\approx$ $-$3.0.

It is important to note also that detailed non-LTE calculations cannot reproduce 
the solar permitted lines (Kiselman \& Nordlund 1995; Kiselman 2001a,b).

(iii) 
Recently, new data on the UV OH lines
(Israelian, Garc\'{\i}a L\'opez \& Rebolo 1998, hereafter IGR98; 
Boesgaard et al. 1999, hereafter BKDV99) 
appear to give high oxygen abundances, 
in agreement with 
the permitted lines, although these lines had been already studied in the 
past by Bessell, Sutherland \& Ruan (1991) and 
Nissen et al. (1994)  who found
[O/Fe] $\approx$ +0.5/+0.6 and +0.6 respectively, in better agreement with the 
forbidden lines. 
It is clear that determinations of oxygen abundances
from UV OH lines are affected however
 at least by different questionable fittings of 
molecular oscillator strengths to the
solar spectrum (see Sect. 4).

(iv) The IR OH transitions were considered as one of the more reliable
oxygen abundance indicators by Grevesse \& Sauval (1994).
Despite this, essentially no oxygen abundance determinations in metal-poor
stars using these lines are available, except for the 
pioneering work by
Balachandran \& Carney (1996, hereafter BC96) for the star HD 103095
for which [Fe/H] = $-$1.22 and [O/Fe] = 0.29  were derived.
Balachandran et al. (2001a,b) presented new oxygen abundance determinations
from IR OH lines; their preliminary [O/Fe] values show
agreement with those derived from [OI] lines.

In the present work we use  infrared OH lines in the region
1.55 - 1.56 $\mu$m, in order to derive oxygen-to-iron ratios
for a sample of 14 metal-poor stars.

In Sect. 2 the observations are presented. In Sect. 3 the detailed
analyses are described.
In Sect. 4 some considerations on UV OH lines are given.
In Sect. 5 the results from IR OH lines are discussed 
and in Sect. 6 conclusions are drawn.

\section{Observations}

\subsection{Data Acquisition}

High-resolution infrared spectra were obtained from images taken  
at the 2.1 m telescope of the Kitt Peak National Observatory 
using the Phoenix spectrograph. Phoenix is described in detail 
by Hinkle et al. (1998). This instrument
uses a cryogenic \'echelle grating to achieve FWHM resolutions of 
50,000 to 75,000, depending on slit width. Order separating filters are 
used to isolate individual \'echelle orders (the instrument is not 
cross dispersed) in the 1 - 5 $\mu$m region. The detector is an 
Aladdin 512x1024 InSb array, covering 1500 km s$^{-1}$ in wavelength.

Each star was observed at least at two slit positions with the same
integration time, with offsets of 15" between integrations,
obtaining two spectra in different positions on the detector. 
The sky background is eliminated by subtracting the first exposure 
from the sky obtained in the second one at that position in the detector, 
and the same method for the second exposure and sky from the first one.
About 30 flatfields and darks were taken, with
exposure times of 15 and 20 s for the H- and K-band, respectively.
Most of the data were gathered in the H-band, centered at 1.5555 $\mu$m, 
with a spectral coverage of 75 {\AA}. This region was chosen because 
it has several unblended vibration-rotation lines of OH, and also 
because it is almost free from disturbing telluric features.
 For a few stars for which no carbon abundance determinations
were available in the literature, we observed the CO lines present at 
2.3 $\mu$m, in order to determine their carbon abundances, since a fraction 
of the oxygen is tied up in CO molecules. 
The observations of OH lines in the H-band
were obtained using the 3-pixel slit, achieving a FWHM resolution 
of 60,000. The K-band observations of  CO lines were done using 
the 4-pixel slit, with a FWHM resolution of 50,000.
For wavelength calibration purposes we have observed the Moon,
obtaining reflected solar spectra. Hot stars with high $v sin i$
were observed to check for the presence of telluric lines
in both  the H and K-band spectra. 
The log of observations is given in Table 1. 

\subsection{Data reduction}

The data were reduced following the standard Phoenix procedures 
{\footnote{ftp://iraf.noao.edu/iraf/misc/phoenix.readme}}, 
with some improvements, in order to obtain resulting spectra of higher quality.

The flatfield and dark frames were combined, and the combined
dark frame was subtracted from the combined flatfield frame.
A mask of bad pixels (both dead and hot pixels)
 was created and these pixels were
reconstructed by interpolation. A response image was created by
fitting a surface to the resulting flatfield.
 The bad pixels were
reconstructed by interpolation using the bad pixel mask. The
spectra were then divided by the response image and one-dimensional 
spectra were extracted (cosmic rays were eliminated in the
process).

\begin{figure}
\epsscale{}
\plotone{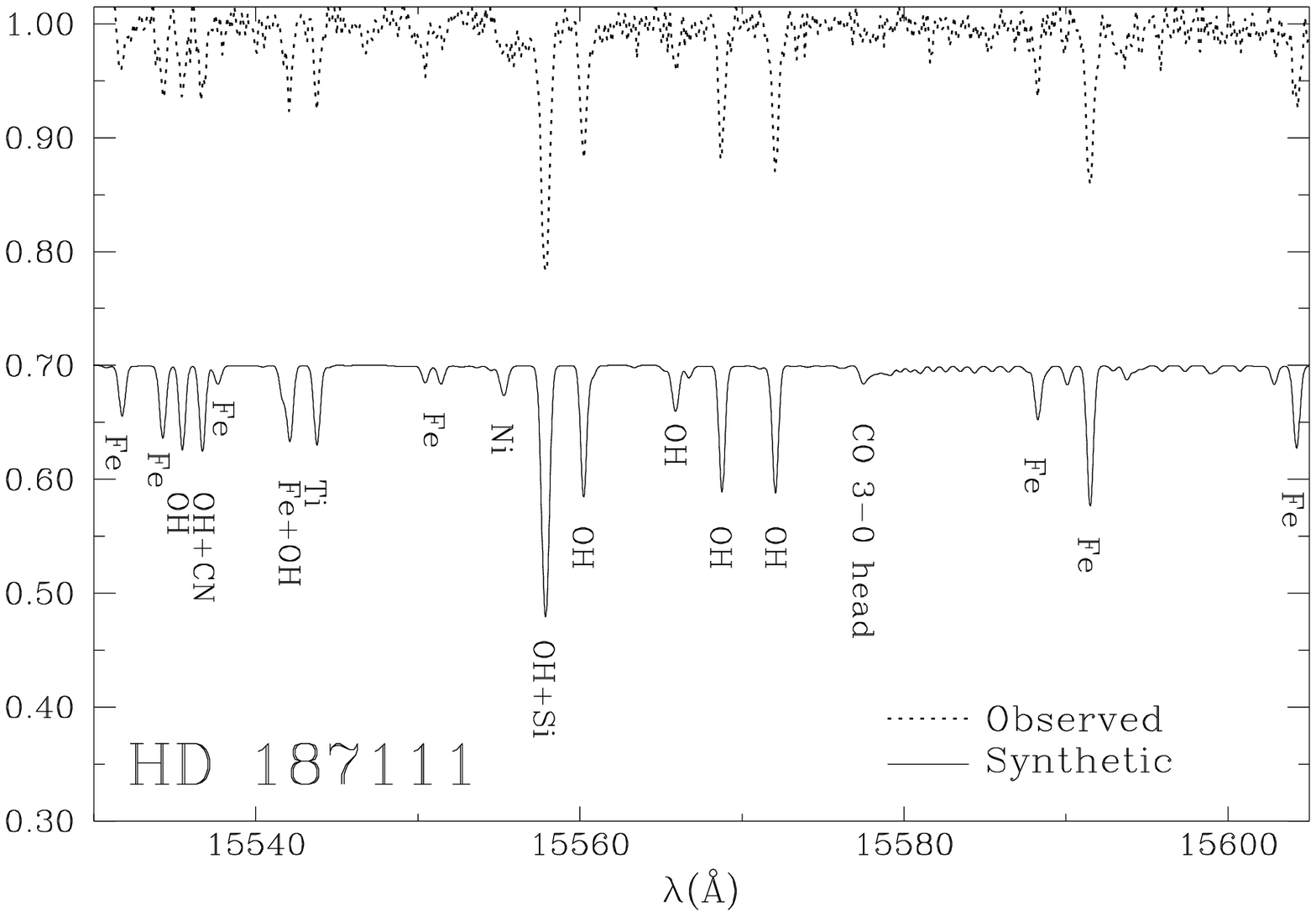}
\caption{PHOENIX normalized spectrum of HD 187111 (upper spectrum) and
synthetic spectrum with line identifications of atomic and molecular
lines (lower spectrum).}
\label{phoenix H band}
\end{figure}

The wavelength calibration of the H-band spectra was done using
the reflected solar spectrum and employing the list of
atomic lines by Mel\'endez \& Barbuy (1999). The solar features were
identified using the solar Atlas of Livingston \& Wallace (1991).
This procedure provided a calibration
with an r.m.s. error of $\Delta \lambda / \lambda \approx 10^{-6}$,
compatible with an error of 0.02 {\AA} in the wavelengths of our
line list. On the first night (Sep. 21st)
the Moon was not observed; instead the giant HR 1481
 of spectral type K1.5III was 
observed (same spectral type of Arcturus) and for spectra 
taken on Sep. 21st we used this giant to carry out
 the wavelength calibration, 
again employing the wavelengths by Mel\'endez \& Barbuy (1999) but 
in this case the identifications were done using the Atlas of Arcturus by 
Hinkle, Wallace \& Livingstone (1995). For comparison purposes we observed
a metal-poor star (HD 37828) in the two nights  and compared the wavelength 
calibrations obtained using either the Moon or the giant spectra. We
found both dispersion solutions in good agreement.
For the K-band spectra we have used telluric features
for wavelength calibration. Telluric spectra were obtained by
observing hot stars with high $v sin i$. Wavelengths of telluric 
lines were measured in the Atlas of Livingston \& Wallace (1991).
We obtained a r.m.s. error of $\Delta \lambda / \lambda \approx 2 
\times 10^{-6}$.
Finally, multiple exposures along the slit were co-added,
and the resulting
spectra have high S/N ratio (column 5 of Table 1), 
typically S/N $\approx$ 100-200, estimated from continuum regions.
The normalized H-band spectrum of HD 187111 is shown in 
Figure \ref{phoenix H band}.

In order to compare our calculations of oxygen abundances 
to those  by BC96, we have extracted the spectrum
of HD 103095 from their publication, using the Dexter utility
{\footnote{http://adsabs.harvard.edu}} (see Sect. 5).

\section{DETAILED ANALYSIS}

The sample consists of 3 dwarfs, 3 subgiants
and 9 giants. They are well-studied metal-poor stars.

\subsection{Effective Temperatures}

Colors available in the literature, used for derivation of
effective temperatures T$_{\rm eff}$, 
were taken from the following sources:
$J$, $H$ and $K$ in the TCS (Telescopio Carlos S\'anchez) system
 from Alonso, 
Arribas \& Mart\'{\i}nez-Roger (1994, 1998); Str\"omgren
 {\it ubvy}-$\beta$ from the Catalogue by Hauck
 \& Mermilliod (1998); $V-I$ (Johnson or Cousins) and
$J$, $H$ and $K$  from the General Catalogue of Photometric Data 
(Mermilliod, Mermilliod \& Hauck 1997). Transformations between
different photometric systems were calculated using relations
given by Bessell (1979), Bessell \& Brett (1988) and 
Alonso et al. (1994, 1998).

The reddening values of most  stars, estimated  by using the
maps of reddening by Burstein \& Heiles (1982), were taken from Anthony-Twarog
\& Twarog (1994). Otherwise
we adopted E(B-V) = 0 for nearby stars (closer than d = 100 pc) and 
 E(B$-$V) = 0.03 csc b [1$-$exp($-$0.008 d(pc) sin $|$b$|$)] (Bond 1980; see also
Arenou et al. 1992) 
for stars with d $>$ 100 pc. Distances were determined from
Hipparcos parallaxes (Perryman et al. 1997). 
The reddening corrections are small, 
typically E(B$-$V) $\approx$ 0.07.
The dereddened colors are given in Table 2.

Temperatures were determined using the $V-I$, $V-K$ and $J-K$
calibrations of Lejeune, Cuisinier \& Buser (1998, LCB98) and 
the same colors plus $b-y$ for the calibrations of 
Alonso, Arribas \& Mart\'{\i}nez-Roger (1996a, 1999a,  
hereafter AAM99a). 
The mean \teff of each calibration is  shown in columns 2 (AAM99a) and 
 3 (LCB98) of Table 3. In a general way, both calibrations 
give similar values, but the calibrations of LCB98 show a larger standard 
deviation of temperatures from different colors 
($\sigma$ $\approx$ 70 K, against $\sigma$ $\approx$  25 K using 
AAM99a).

In Table 3 are also shown Infrared Flux Method (IRFM, column 4) temperatures,
as determined by Alonso, Arribas \& Mart\'{\i}nez-Roger (1996b, 1999b, 
hereafter AAM99b). 
The calibration of AAM99a is based on their own IRFM temperature 
determinations, 
which explains  the close values given under AAM99a and IRFM (AAM99b)
 in Table 3. 
The calibration of LCB98 is slightly cooler ($-$10 K in the mean) than IRFM 
temperatures.

Means of T$_{\rm eff}$ values derived from colors (columns 2, 3 and 4
of Table 3) were  checked against
excitation equilibrium of \ion{Fe}{1} lines. 
Literature equivalent widths of \ion{Fe}{1} and \ion{Fe}{2} (references
in Sect. 3.3) were used.

The temperatures based on excitation equilibrium of \ion{Fe}{1}  are
in the mean 40 K hotter than the mean of 
T$_{\rm eff}$ values derived from colors, and
+50 K hotter in the mean relative to the IRFM temperatures
 (or +38 K if we consider the IRFM calibration of AAM99a). 
Note that Tomkin \& Lambert (1999) found a similar result, with their \ion{Fe}{1} 
temperatures being +45 K higher than those determined from the IRFM calibration of 
AAM99a.  According to Tomkin \& Lambert (1999), the small difference between 
\ion{Fe}{1} temperatures and the IRFM temperatures is an indication that 
non-LTE effects are minor.

\begin{figure}
\epsscale{}
\plotone{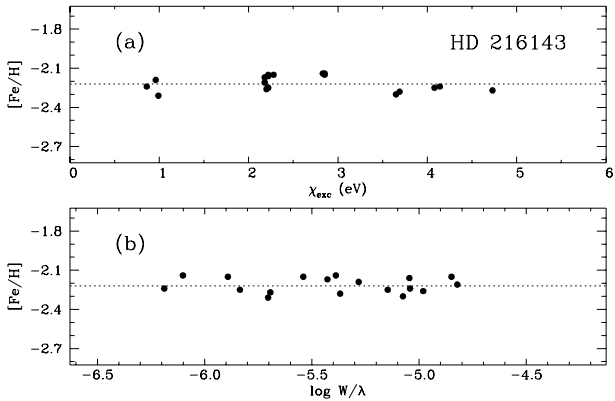}
\caption{[Fe/H] vs. (a) excitation potential $\chi_{exc}$ and
(b) reduced equivalent width W/$\protect\lambda$
 for the spectroscopic parameters of HD 216143. 
There is no significant trend with $\protect\chi_{exc}$ or log W/$\protect\lambda$.}
\label{T4344}
\end{figure}

An example of the excitation equilibrium for HD 216143 is shown 
in Figs. \ref{T4344}a,b, 
where the individual iron abundances are plotted 
versus excitation potential and reduced equivalent width.
 No significant slope is seen in these Figs., indicating 
that \teff = 4344 K and v$_t$ = 1.8 km s$^{-1}$ are 
adequate choices.

Two sets of  T$_{\rm eff}$ values were adopted, those 
resulting from the excitation equilibrium (column 5 of Table 3)
and those based on IRFM (column 4 of Table 3).

\subsection{Gravities}

Nissen et al. (1997) and  Allende Prieto et al. (1999)
have shown that LTE spectroscopic gravities are sistematically
lower than trigonometric gravities derived from Hipparcos
parallaxes in metal-poor stars.
For this reason, both spectroscopic and trigonometric gravities
were derived.

Spectroscopic gravities obtained by requiring ionization equilibrium of 
\ion{Fe}{1} and \ion{Fe}{2} lines
are given in Column 3 of Table 4.
Hipparcos parallaxes $\pi$ are used to derive trigonometric gravities,
given in Column 7 of Table 4, where 
are also given the errors corresponding to the standard deviation
$\sigma (\pi )$, which is of the order of 1 to 2 mas (a mean
around 50\% in $\pi$). The errors in the trigonometric log $g$
are large for giants with log $g$ $\simless$ 1.5, and small
for dwarfs, in which case the gravities
 are also close to the spectroscopic values.

\subsection{Metallicities}

Equivalent widths of \ion{Fe}{1} and \ion{Fe}{2} 
 from recent high resolution work in the literature 
(Kraft et al. 1992; Tomkin et al. 1992; Beveridge \& Sneden 1994; 
Roe, Pilachowski \& Armandroff 1994; Pilachowski,
Sneden \& Kraft 1996; Shetrone 1996; Tomkin \& Lambert 1999; Fulbright 2000) 
were adopted. Optical spectra by Barbuy (1988) and 
Barbuy \& Erdelyi-Mendes (1989) were used to check for any 
systematic error in the adopted equivalent widths.
The present infrared spectra of the program stars, which contained
a number of \ion{Fe}{1}  lines, were also used for a further
check on [Fe/H]. The IR \ion{Fe}{1} lines and respective 
equivalent widths are reported in Table 5.

MARCS model atmospheres by Bell et al. (1976) and Gustafsson et al. (1975)
and OSMARCS by Edvardsson et al. (1993) were used for the 
calculations of curves of growth and spectrum synthesis.
OSMARCS models were applied to 3 dwarf stars.
The differences in [Fe/H] with respect to MARCS
were $\Delta$[Fe/H](OSMARCS-MARCS) $\leq$ 0.05. 
For the sake of consistency with the other sample stars,
we assumed the results from MARCS models.

The adopted oscillator strengths for the optical iron lines are from 
the National Institute of Standards \& Technology 
 (Fuhr, Martin \& Wiese 1988). For the infrared atomic lines we 
used the solar oscillator strengths determined in Mel\'endez \& Barbuy 
(1999), which were derived adopting solar abundances from Grevesse, 
Noels \& Sauval (1996). Curves of growth and abundances 
from equivalent widths of \ion{Fe}{1} and \ion{Fe}{2} 
were computed using the codes RENOIR and ABON by M. Spite. 

In Table 7a are given the Fe abundances derived from (a) curves
of growth of optical \ion{Fe}{1} lines and checked with
 IR \ion{Fe}{1} lines, obtained by adopting spectroscopic effective
temperatures and gravities (columns 2 and 3 of Table 7a), (b) IRFM 
effective temperatures, Hipparcos gravities, and curves of
growth based on optical \ion{Fe}{2} lines, and checked against
the IR \ion{Fe}{1} lines (columns 4 and 5 of Table 7a).
Note that the agreement of the optical and IR-based Fe abundances
shows that: a) there are no problems with the
Phoenix data (such as scattered light) that might affect the equivalent
widths which could cause the IR OH abundances to be low with
respect to the UV OH, and b) there is no opacity problem, or
other errors in the spectral code, between the optical and 
infrared regions.

\begin{figure}
\plotone{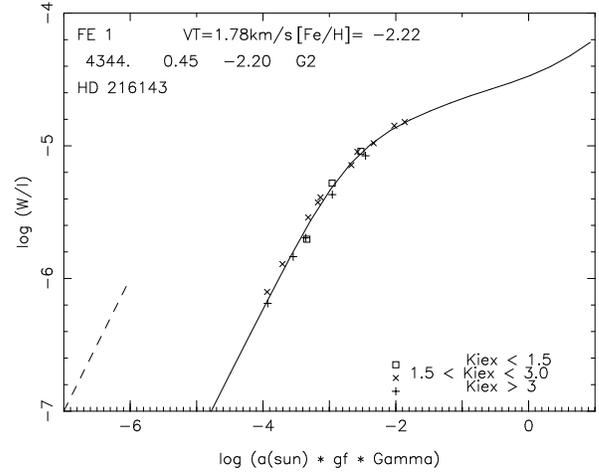}
\caption{Curve of growth of \protect\ion{Fe}{1} for HD 216143.}
\label{cdc 216143}
\end{figure}

Microturbulence velocities v$_t$ were obtained from  curves of growth
and these values were checked by requiring no dependence 
of [Fe/H] against reduced equivalent width. 
The curve of growth for HD 216143, 
the most metal-poor star of the present sample,
is shown in Figure \ref{cdc 216143}.

The two sets of stellar parameters derived, and values from the
literature are given in Table 4, where columns 2-5 show the spectroscopic
parameters, co\-lumns 6-9 show the IRFM T$_{\rm eff}$s,
trigonometric parallaxes and
[\ion{Fe}{2}/H], and literature values are given in columns 10-13.

\subsection{Oxygen Abundances}

The oxygen abundances were determined from fits of synthetic spectra
to the sample spectra. The LTE code for spectrum synthesis 
described in  Cayrel et al. (1991)
was employed for the calculations.

The list of  atomic lines present in the H band 
 compiled by Mel\'endez \& Barbuy (1999)
was adopted, where oscillator strengths and damping constants were 
obtained from a line-by-line fitting of $\approx$ 2,200 lines in the 
infrared $J$ and $H$ bands, using the telluric-free solar spectrum atlas of 
Livingston \& Wallace (1991) and Wallace, Hinkle \& Livingston (1993). 
Hyperfine structure was also taken into account, as described in 
Mel\'endez (1999). Molecular lines of CN A$^2$$\Pi$ - X$^2$$\Sigma$, 
CO X$^1$$\Sigma^+$  (Mel\'endez \& Barbuy 1999) and OH (X$^2$$\Pi$) are included.

The infrared OH vibration-rotation lines (X$^2 \Pi$) used in this work are
from the first-overtone ($\Delta$ v = 2) sequence. 
Laboratory line lists with  wavenumbers and identifications
were kindly provided by S. P. Davis. These line lists are from the work
of Abrams et al. (1994). A few lines identified by M\'elen et al. (1995), 
 and a few theoretical lines (replacing unidentified lines) from 
Goldman et al. (1998) were also included.
Energy levels for the OH lines were computed from molecular parameters 
given in Coxon \& Foster (1992) and Abrams et al. (1994). 
Molecular oscillator strengths were calculated from 
Einstein coefficients given in the extensive work by Goldman et al. (1998).
A complete electronic version of line parameters for OH X$^2 \Pi$ 
transitions was kindly made available to us by A. Goldman. 
To our knowledge, Goldman et al. (1998) provide
the most accurate Einstein values for infrared OH transitions
available in the literature. The adopted dissociation potential
is 4.392 eV (Huber \& Herzberg 1979).

The main atomic and molecular features of the observed region are shown 
in Figure \ref{phoenix H band}.
The list of OH lines used for oxygen abundance determination,
together with their molecular gf-values, energy levels and
 equivalent widths are given in Table 5.

The oxygen abundances derived show a dependence on
 the carbon abundances adopted.
For 7 stars, carbon abundances were determined from the 
CO 3-0 bandhead (1.56 $\mu$m) or from CO lines of the 
2-0 and 3-1 bands (2.33 $\mu$m), otherwise [C/Fe] values were taken from the 
literature. [C/Fe] values obtained from spectrum synthesis of the 
infrared CO bands are given in column 2 of Table 6. 
 In the last column are given values from the 
literature, where most results are based on CH lines, except for
the BC96 value for HD 103095 based on IR CO lines.
The carbon abundances obtained from CO lines in the present work
 show good agreement with  literature data.

\begin{figure}
\epsscale{}
\plotone{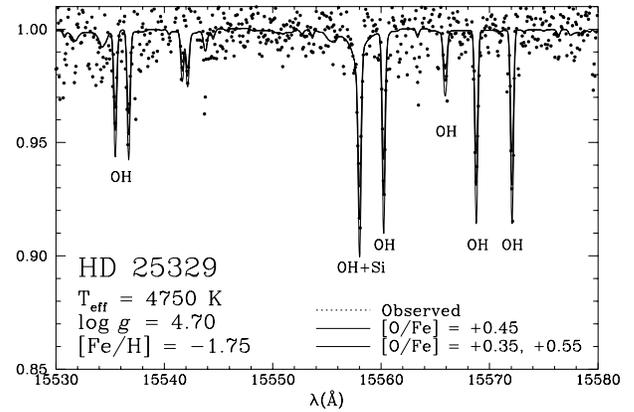}
\caption{Spectrum of HD 25329 (dots) compared to 
synthetic spectra computed with  
[O/Fe]: +0.35, +0.45 (thick line) and +0.55.}
\label{oh 25329}
\end{figure}

\begin{figure}
\epsscale{}
\plotone{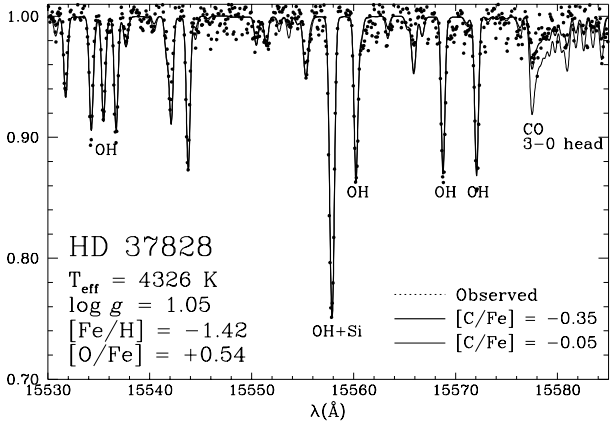}
\caption{Observed spectrum of HD 37828 (dots) compared to
synthetic spectra computed  with
[O/Fe] = +0.54,  [C/Fe] = $-$0.35 (thick solid line)
 and  [C/Fe] = $-$0.05 (thin solid line).}
\label{oh 37828}
\end{figure}

Figures \ref{oh 25329} and \ref{oh 37828} show
the fit of synthetic spectra to the observed spectra of
HD 25329, the most metal-poor dwarf of our sample, 
and HD 37828, where the behavior of CO bands is illustrated.

\begin{figure}
\epsscale{}
\plotone{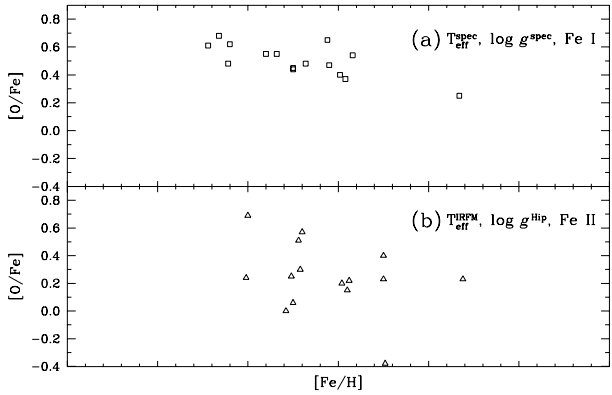}
\caption{[O/Fe] vs. [Fe/H] from IR OH lines.
(a) open squares correspond to results based on spectroscopic parameters;
(b) open triangles give results from stellar parameters based on IRFM 
temperatures, trigonometric gravities
and [Fe/H] derived from \protect\ion{Fe}{2} lines. 
}
\label{ofe hipspec}
\end{figure}

The oxygen abundances derived from 
(a) spectroscopic parameters and (b) IRFM T$_{\rm eff}$s, trigonometric
gravities and [\ion{Fe}{2}/H] are given in Table 7b.
In Figures \ref{ofe hipspec}a,b are plotted [O/Fe] vs. [Fe/H] 
obtained from (a) spectroscopic parameters (open squares) and
(b) IRFM temperatures, trigonometric gravities
and [Fe/H] derived from \ion{Fe}{2} lines (open triangles).
Note that the uncertainty on trigonometric gravities cause
a large dispersion on oxygen-to-iron ratios.
We adopted as final result, the mean of methods (a) and (b), 
given in column 10 of Table 7b.

Adopting the spectroscopic parameters a mean 
oxygen-to-iron ratio of [O/Fe] $\approx$ +0.52 is obtained, 
whereas using the IRFM temperatures, Hipparcos gravities 
and [\ion{Fe}{2}/H], [O/Fe] $\approx$ +0.25 is found.
A mean of the two methods gives a final 
value of [O/Fe] $\approx$  +0.38$\pm$0.2 
for the metallicity range of the sample metal-poor stars 
of $-$2.2 $<$ [Fe/H] $<$ $-$1.2. Note that we have used Grevesse et al.
(1996) abundance values, where $\epsilon$(O) = 8.87. If
the Grevesse \& Sauval (1998) $\epsilon$(O) = 8.83 were used,
[O/Fe] values would be higher by + 0.04 dex.
The present oxygen abundances are in agreement with results from [\ion{O}{1}]
forbidden lines, which applies also to the 3 dwarfs of the present sample.

\subsection{Errors}

A check on the [Fe/H] values and possible systematic errors 
in the adopted equivalent widths was 
carried out by employing the high resolution 
optical spectra by Barbuy (1988) and 
Barbuy \& Erdelyi-Mendes (1989). For five stars in common 
with the Lick, Texas and KPNO data,
 the \ion{Fe}{1} equivalent widths W$_{\lambda}$ were found to be
about 10\% smaller than those adopted; this 
difference is negligible since it implies a change
in [Fe/H] of only 0.04 dex.
A further check on [Fe/H] values was done with the infrared spectra 
observed in the present work, since five non-blended and other blended
high excitation (5.5 to 6.5 eV) \ion{Fe}{1} lines are 
observed in the region.
The [Fe/H]$_{\rm IR}$  abundances  determined 
by computing synthetic spectra (columns 3 and 5 of Table 7a) 
are in good agreement with the iron abundances 
from the analyses  (columns 2 and 4  of Table 7a),
within $\Delta$[Fe/H] $\leq$ 0.15, and
a mean difference $\Delta$[Fe/H] $\approx$ +0.06
and $\Delta$[Fe/H] $\approx$ $-$0.15  for 
parameters (a) and (b) (Sect. 3.3) respectively.

The errors due to uncertainties on T$_{\rm eff}$, log $g$
and v$_{\rm t}$ are inspected by computing the results for
3 sample stars. In Table 8 a mean of the errors for
the giants HD 37828 and HD 216143, and those for the
dwarf HD 103095 are given, where $\Delta$T$_{\rm eff}$ = 100 K,
$\Delta$log $g$ = 0.5 dex and $\Delta$v$_{\rm t}$ = 0.5 km s$^{-1}$
are assumed.

\section{Comments on UV OH lines}

Using UV OH lines, both IGR98 and
BKDV99 obtained high oxygen abundances in
metal-poor stars, which appear to be in agreement with most
results from permitted lines  (but see Carretta et al. 2000).
As already noted in Sect. 1,  previous work on the UV OH 
lines resulted in lower oxygen
abundances of [O/Fe] $\approx$ +0.5 to +0.6
for metal-poor stars (Bessell, Sutherland \& Ruan 1991;
Nissen et al. 1994) in better agreement with the 
forbidden lines.

Let us analyse the possible sources of uncertainties in
the use of the UV OH lines:

Nissen et al. (1994) transformed the Einstein $A$ coefficients
for the (0,0) transition calculated by Goldman \& Gillis (1981)
to molecular gf-values. In order to compute the same quantity
for the (1,1) transition, Franck-Condon factors by Felenbok (1963)
were adopted.
Gillis et al. (2001) find results very similar to 
Goldman \& Gillis (1981),
and now include results for several (v'v") vibrational transitions.
In order to fit the OH lines in the solar spectrum, 
Nissen et al. (1994) corrected
the molecular gf-values by
$-$0.16 for the (v'v") = (0,0) and $-$0.49 dex for the (1,1) vibrational
transition.
Calculations of Franck-Condon factors using the code by Jarmain \& 
McCallum (1970) for the UV OH transitions (P.D. Singh, 
private communication), give  values 
of q$_{00}$ = 0.864 and q$_{11}$ =  0.683, or  q$_{11}$/q$_{00}$ = 0.79, 
the same ratio found from Felenbok (1963) results, 
whereas the ratio of molecular oscillator strengths
gf(11)/gf(00) = 0.63 following the new Einstein $A$
coefficients by Gillis et al. (2001).
With the corrections by Nissen et al. (1994), also adopted by 
BKDV99, this ratio becomes gf(11)/gf(00) = 0.37, much lower
than that given in calculations of Franck-Condon factors.

\begin{figure}
\epsscale{}
\plotone{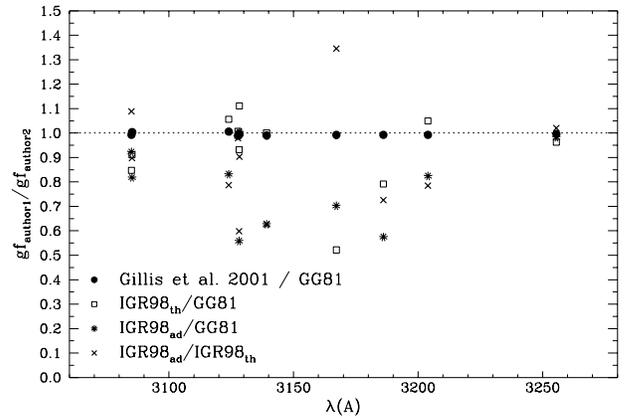}
\caption{
Ratios of  molecular gf-values from different authors on a 
line-by-line basis: (a) Gillis et al. (2001) relative
to Goldman \& Gillis (1981), which are essentially the same 
(filled circles);
(b) theoretical values computed by IGR98 relative
to Goldman \& Gillis (1981), where unexpected differences are seen
(open squares);
(c) IGR98 values fits to the solar spectrum relative to
Goldman \& Gillis (1981), where IGR98 values are systematically
lower (asterisks); (d) IGR98 adopted relative to
 theoretical values according to IGR98 (crosses).}
\label{gf oh}
\end{figure}

IGR98 applied corrections on a OH line-by-line basis, 
in order to match the observed OH solar lines.
Differences on gf-values employed by them, relative to Gillis
et al. (2001) are down to -0.25 dex (represented by
asterisks in Fig. \ref{gf oh}).
In order to understand which may be the errors in IGR98's
molecular gf-values, we computed the gf-values based on
Goldman \& Gillis (1981) and Gillis et al. (2001) and compared
these values to IGR98's ones. 
In Figure \ref{gf oh} are plotted ratios of line-by-line molecular gf-values
between different authors: (a) Gillis et al. (2001) relative
to Goldman \& Gillis (1981), which are essentially the same (filled circles);
(b) theoretical values computed by IGR98 relative
to Goldman \& Gillis (1981), where unexpected differences are seen
(open squares);
(c) IGR98 values matched to the solar spectrum relative to
Goldman \& Gillis (1981), where IGR98 values are systematically
lower (asterisks); (d) IGR98 adopted relative to theoretical values according
to those authors (crosses).
The IGR98 oxygen abundances would be lower by 0.1-0.2 dex if
the original theoretical gf-values were used. 
Also in the sense of lowering further the oxygen abundances
derived by IGR98, Asplund (2001a,b) suggested that calculations 
of UV OH lines would be more appropriate by taking into account 
3-D model atmospheres, since these lines are very sensitive to 
temperature variations. 

\section{Discussion}

A comparison of the present atmospheric parameters with literature values
is given in Table 4 and oxygen abundances available from [OI],
OI, UV OH and IR OH lines for the sample stars are reported in Table 9.

{\it [OI] lines}:
The oxygen abundances obtained in the present work
are in agreement with those derived from the [OI] line, and in
a general agreement with values based on the forbidden 
[OI] lines given in the literature, which give a mean of [O/Fe]
$\approx$ 0.4 to 0.6 in the metallicity range
$-$2.5 $<$ [Fe/H] $<$ $-$1.2 (see references in Sect. 1).

{\it OI triplet lines:}
A discrepancy is seen between the presently derived O abundances
from IR OH lines, together with those from the [OI] lines,
 with respect to those based on the permitted
OI triplet lines. In order to try to solve this 
well-known discrepancy, King (1993) and Carretta et al. (2000)
adopted higher effective temperature scales, which
 lower the [O/Fe] ratios deduced from 
the \ion{O}{1} triplet bringing these into agreement with the
values obtained from the [\ion{O}{1}] lines.
For  HD 103095 the effective temperature of 
T$_{\rm eff}$ = 5132 K adopted by
BKDV99, in the King scale,
 bring all oxygen abundance indicators into agreement. 
It implies an
increase of +100 K in \teff relative to the temperatures
obtained from both \ion{Fe}{1} and the IRFM.
Note that recent high resolution spectroscopic analyses 
of HD 103095 by BC96, 
Tomkin \& Lambert (1999) and Fulbright (2000) present temperatures 
of \teff $\approx$ 5000 $\pm$ 50 K, in
agreement with recent IRFM estimates
of \teff = 5029 K (Alonso et al. 1996b). 
While an increase of +100 K is required for this particular case, 
in other cases temperature differences of up to +400 K are required to 
bring the oxygen abundances derived from the \ion{O}{1} triplet into 
agreement with 
the values obtained from [\ion{O}{1}] lines (see Cavallo et al. 1997).

\begin{figure}
\epsscale{}
\plotone{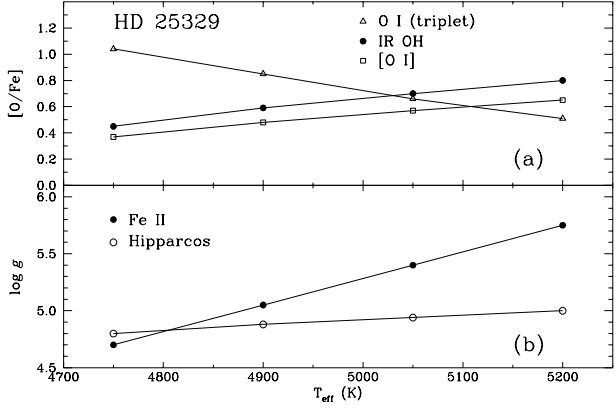}
\protect\caption{
a) [O/Fe] vs. \protect\teff and log {\protect\it g} for HD 25329.
\protect\teff (and correspondingly log {\protect\it g}) has
to be increased to bring the oxygen abundance from the IR OH lines
(filled circles) and [\protect\ion{O}{1}] (open squares) into agreement with
the abundance derived from the \protect\ion{O}{1} triplet (open triangles).
For [\protect\ion{O}{1}] and \protect\ion{O}{1} the abundances were obtained from data 
by Spiesman \& Wallerstein (1991) and Beveridge \& Sneden (1994), respectively.
b) Log {\protect\it g} vs. \protect\teff for HD 25329 obtained from
ionization equilibrium (filled circles) and Hipparcos parallaxes (open circles).
Note that for \protect\teff $\approx$ 4750 - 4800 K (as obtained from \protect\ion{Fe}{1} and
the IRFM) there is agreement between the gravities obtained from both methods.
}
\label{25329 test}
\end{figure}

In Figure \ref{25329 test}a  the [O/Fe] ratios obtained from 
the IR OH lines, [\ion{O}{1}] and \ion{O}{1} for the star HD 25329 are shown
for a range of temperatures: the \teff has to be raised by
$\Delta$T$_{\rm eff}$ $\approx$ +300/+350 K,
well above the most likely values, in order to find an agreement
(or even higher  if Hipparcos gravities are used), 
 relative to the spectroscopic value.
The gravity obtained by ionization equilibrium from such 
an increase of 300 K in \teff for 
HD 25329,  is of log {\it g} $\approx$  5.5 dex, which is
in disagreement by $\Delta$ log {\it g} $\approx $+0.6 dex  
with respect to the log $g$ value obtained from its Hipparcos parallax, 
as shown in Figure \ref {25329 test}b.

Other problems with these lines need further investigation:
the O abundances from the \ion{O}{1} triplet lines show 
a large scatter (see Fig. 4 of Mishenina et al. 2000
and Fig. 7 of Cavallo et al. 1997)
and some \teff dependence (see Fig. 8 of Tomkin et al. 1992
and Fig. 2 of Carretta et al. 2000).
See further discussion on these lines in
Cayrel (2001b), Kiselman (2001a,b), Asplund (2001a,b)
and Lambert (2001).

{\it UV OH lines:}
Only two sample stars had the UV OH lines analysed: for HD 6582
there is good agreement between the present IR OH values and
 IGR98. For HD 103095  differences of
$\Delta$[O/Fe] = $-$0.32 with respect to IGR98, and $-$0.14 (King
scale) and 0.0 (Carney scale)
with respect to BKDV99 are found; these differences
are small, and if the gf-values of IGR98 and BKDV99
were assumed to be the original ones by Gillis et al. (2001), 
essentially no discrepancy would appear.
Therefore no discrepancy between oxygen abundances derived from UV OH and
IR OH are seen in these two dwarfs. 
However, considering the entire samples of UV OH results by BKDV99, 
Boesgaard (2001a,b), Garc\'{\i}a L\'opez
et al. (2001a,b) and IGR98 on one hand,
and IR OH results by Balachandran et al. (2001a,b) 
and the present results (see also Mel\'endez et al. 2001)
on the other,  a sizeable discrepancy is seen
of the order of $\Delta$[O/Fe] $\sim$ 0.5 at [Fe/H] $\sim$ $-$2.0,
and still undefined for lower metallicities.

\begin{figure}
\epsscale{}
\plotone{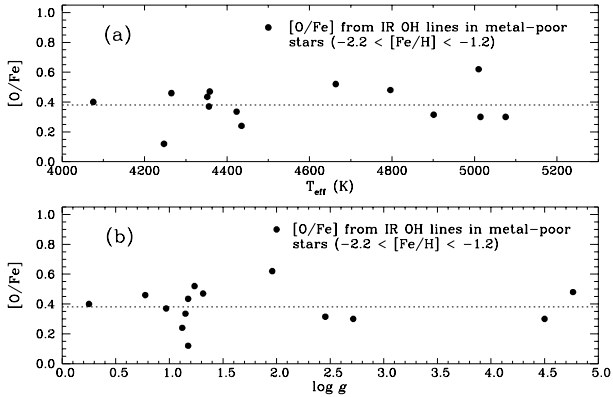}
\caption{[O/Fe] from IR OH lines vs. (a) \protect\teff
and (b) log $g$, for stars with [Fe/H] $<$ $-$1.2.}

\label{ofeteff}
\end{figure}

{\it IR OH lines:}
 [O/Fe] values derived from the IR OH lines as a function of 
\teff are plotted in Fig. \ref{ofeteff}a, where neither any
significant scatter nor \teff dependence are seen.
Oxygen abundance determinations from IR OH lines in the literature
were available up to now only for the metal-poor star HD 103095 by
BC96. They obtained [O/Fe] = +0.29 dex 
for the halo dwarf HD 103095, in good agreement with the present result of
[O/Fe] = +0.30 (or +0.28 if BC96 parameters are used).
In Table 10 are given [O/Fe] values derived from the IR OH lines
for HD 103095, by adopting six different model parameters from 
the literature, showing that the [O/Fe] value obtained  is not 
very sensitive to the model atmosphere parameters. 
Balachandran et al. (2001a,b) obtained [O/Fe] $\approx$ +0.4 for
6 stars with metallicities $-$2.6 $\leq$ [Fe/H] $\leq$ $-$1.0, in agreement
with the present work. For 3 stars with [Fe/H] = $-$2.6, $-$2.7 and $-$3.0,
they find [O/Fe] = +0.7, +0.7 and +0.9 respectively. It is also
important to note that for the star BD+23 3130, their results are
in agreeement with UV OH results from IGR98 if the original
molecular gf-values by Goldman \& Gillis (1981) are used,
and with that obtained from the
[OI] 6300 {\rm \AA} line by Cayrel et al. (2001a).  

In Figure \ref{ofeteff}b  the [O/Fe] values 
(in stars) of [Fe/H] $<$ $-$1.2 vs. gravity are plotted.
Oxygen abundances do not seem depleted in these metal-poor field stars.
This is not unexpected since,
in a recent work, Gratton et al. (2000) studied the mixing along the red giant
branch in 62 metal-poor field stars, where 
none of their more evolved {\it field} giants (about half of the sample) shows 
any sign of oxygen depletion.

\begin{figure}
\epsscale{}
\plotone{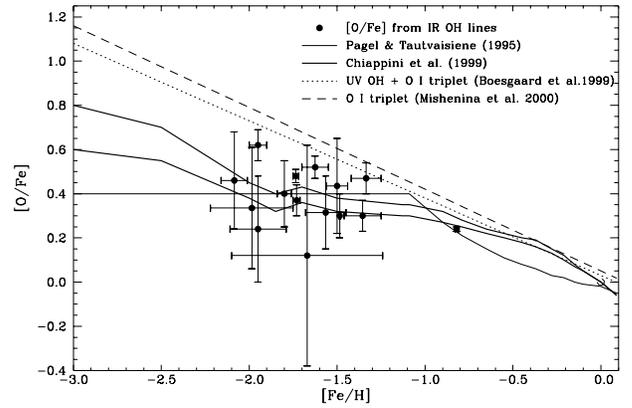}
\protect\caption{
Comparison of oxygen abundances
from IR OH lines (filled circles) with chemical evolution models. 
Fits of the average of the UV OH  and OI triplet abundances by
Boesgaard et al. (1999) (dotted line) and of the OI triplet abundances
by Mishenina et al. (2000) (dashed line) are shown.
Predictions of oxygen abundances from chemical evolution models
by Chiappini et al. (1999) (thick line) and 
by Pagel \& Tautvai\v{s}ien\.{e} (1995) (thin line) are also plotted. 
The extremes of the error bars correspond to the (i) spectroscopic parameters 
and (ii) assuming IRFM T$_{\protect\rm eff}$s, trigonometric gravities 
and [\protect\ion{Fe}{2}/H] (see Sect. 3).
The open circle represents the Sun.
}
\label{ofe final}
\end{figure}

In Figure \ref{ofe final} are shown the present [O/Fe] determinations 
compared with the chemical evolution models of CMBN99 and PT95. 
The former model  presents predictions of abundances relative to
iron for an extended range of metallicities ($-$4.0 $<$ [Fe/H] $<$ 0.0).
No instantaneous recyling is assumed, and two different sets of yields
from massive stars, one from WW95 and the other from TNH96
are employed. Their
[O/Fe] values show a slope with decreasing metallicity, which is more
pronounced if TNH96 yields are assumed. Similar predictions are given
for other $\alpha$ elements such as Mg, S, Si, Ca, for which the WW95
yields seem more compatible with the observational data.
For the present range of metallicities, both their 
models reproduce reasonably well  the observed [O/Fe] ratios.
 It would be important to 
test the prediction of increasing [O/Fe] with decreasing metallicity
 by measuring IR OH lines  in stars of lower metallicities ([Fe/H] $<$ $-$2.5).
The PT95 models present two versions, assuming instantaneous or
 delayed production approximations,
using both TNH96 and WW95 yields, and their mean value for the range of
metallicities $-$2.5 $<$ [Fe/H] $<$ 0.0 is consistent with the present data.

\section{Conclusions}

We obtained high-resolution infrared spectra in the H-band,
in order to derive oxygen abundances from IR OH lines.
In order to have a homegeneous set of stellar parameters
we carried out detailed analyses using  
equivalent widths of iron lines from the Lick, Texas and KPNO groups.

The stellar parameters were derived with basis on two methods:
(a) spectroscopic parameters and (b) effective temperatures derived
from the Infrared Flux Method (IRFM), trigonometric gravities using
Hipparcos parallaxes, and metallicities [Fe/H] based on curves
of growth of \ion{Fe}{2}.
Using stellar parameters of set (a), a mean [O/Fe] = +0.52 is
found, with a possible slope giving higher oxygen-to-iron ratios
with decreasing metallicities, whereas with set (b) a constant
[O/Fe] = +0.25 is obtained. A mean of the two methods gives
[O/Fe] = +0.38 (assuming $\epsilon$(O) = 8.87 and $\epsilon$(Fe) = 7.50).

In summary,  the sample stars with metallicities in the range 
$-$2.2 $<$ [Fe/H] $<$ $-$1.2
show  [O/Fe] $\approx$ 0.4$\pm$0.2, with no significant evidence 
for an increase of [O/Fe] with decreasing metallicity.

\acknowledgements
We are grateful to the PHOENIX team
at Kitt Peak, especially to  the instrument scientist
Ken Hinkle. We thank J. Tomkin, C. Pilachowski
and J. Fulbright for providing 
unpublished equivalent widths, and S. P. Davis and A. Goldman for 
sending electronic files with OH molecular data. 
We are also grateful to a very competent referee, who helped us
to improve the original manuscript.
We acknowledge partial financial support from FAPESP, CNPq and CNPq/CNRS.
J.M. acknowledges the FAPESP PhD fellowship n$^{\rm o}$ 97/00109-8.
We  have made use of data from the Hipparcos astrometric mission of
the ESA.

\end{document}


\newcommand{\teff}{T$_{\rm eff}$ }

\begin{deluxetable}{llllll}
\tablenum{1}
\label{observations}
\tablewidth{0pt}
\footnotesize
\tablecaption{Log of observations}
\tablehead{
\colhead{Star}  &
\colhead{V}      &
\colhead{Date (Sep. 1999)}      &
\colhead{Exposure (s)}      &
\colhead{S/N} &
\colhead{band} }
\startdata
HD 2665    & 7.7 & 26 & 3x1500 & 110 & H \\
HD 6582    & 5.1 & 25, 27 & 3x600, 2x300 & 180, 100 & H, K \\
HD 6755    & 7.7 & 25 & 4x1200 & 200 & H \\
HD 21581   & 8.7 & 27 & 4x1800 & 100 & H \\
HD 25329   & 8.5 & 25 & 6x1800 & 150 & H \\
HD 26297   & 7.5 & 21 & 2x1200 & 130 & H \\
HD 29574   & 8.3 & 21 & 2x1000 & 80  & H \\
HD 37828   & 6.9 & 21, 26, 27  & 2x300, 2x300, 2x300 & 
                                105, 105, 100 & H, H, K \\
HD 165195  & 7.3 & 26 & 2x1500 & 200 & H \\
HD 187111  & 7.7 & 26, 27 & 2x900,  2x1200 & 140, 150 & H, K \\
HD 206739  & 8.6 & 27 & 3x1800 & 120 & H \\
HD 216143  & 7.8 & 26 & 4x1800 & 230 & H \\
HD 221170  & 7.7 & 25 & 4x1500 & 180 & H \\
BD +060648 & 9.1 & 26 & 4x1800 & 180 & H \\
\enddata
\end{deluxetable}

\begin{deluxetable}{llllllll}
\tablenum{2}
\label{colors}
\tablewidth{0pt}
\footnotesize
\tablecaption{Colors}
\tablehead{
\colhead{Star} &
\colhead{{(b-y)}$_{\rm 0}$} &
\colhead{m$_{\rm 0}$}      &
\colhead{c$_{\rm 0}$}      &
\colhead{{(V-I)}$^{\rm C}_{\rm0}$} &
\colhead{{(V-I)}$^{\rm J}_{\rm 0}$} &
\colhead{{(J-K)}$^{\rm TCS}_{\rm 0}$} &
\colhead{{(V-K)}$^{\rm TCS}_{\rm 0}$} }
\startdata
HD 2665   & 0.502 & 0.067 & 0.340 & 1.166 & 1.500 & 0.530 & 2.092 \\
HD 6582   & 0.437 & 0.205 & 0.213 &  --   &  --   & 0.443 & 1.833 \\
HD 6755   & 0.474 & 0.095 & 0.290 & 0.854 & 1.097 & 0.518 & 1.988 \\
HD 21581  & 0.527 & 0.117 & 0.331 & 0.943 & 1.212 & 0.546 & 2.189 \\
HD 25329  & 0.528 & 0.303 & 0.155 & 0.988 & 1.270 & 0.557 & 2.305 \\
HD 26297  & 0.738 & 0.237 & 0.607 &  --   &  --   & 0.713 & 2.845 \\
HD 29574  & 0.931 & 0.219 & 0.721 &  --   &  --   & 0.806 & 3.368 \\
HD 37828  & 0.694 & 0.308 & 0.491 & 1.190 & 1.530 &   --  &  --   \\
HD 103095 & 0.484 & 0.222 & 0.155 & 0.871 & 1.120 & 0.492 & 2.054 \\
HD 165195 & 0.821 & 0.120 & 0.701 & 1.286 & 1.653 & 0.749 & 2.936 \\
HD 187111 & 0.754 & 0.197 & 0.588 &  --   &  --   & 0.770 & 2.917 \\
HD 206739 & 0.602 & 0.172 & 0.446 &  --   &  --   & 0.603 & 2.413 \\
HD 216143 & 0.677 & 0.150 & 0.569 &  --   &  --   & 0.653 & 2.617 \\
HD 221170 & 0.713 & 0.114 & 0.547 &  --   &  --   & 0.685 & 2.692 \\
BD +060648 & 0.787 & 0.175 & 0.577 & --   &  --   & 0.740 & 2.932 \\
\enddata
\tablecomments{C = Cousins, J = Johnson, TCS = Telescopio Carlos S\'anchez}
\end{deluxetable}

\begin{deluxetable}{llllll}
\tablenum{3}
\tablewidth{0pt}
\footnotesize
\tablecaption{Effective temperatures (K) based on different methods (Sect. 3.1)}
\tablehead{
\colhead{Star}   &
\colhead{AAM99a} &
\colhead{LCB98}  &
\colhead{IRFM} &
\colhead{\ion{Fe}{1}} }
\startdata
HD 2665   & 4964 & 4980 & 4990 &  5030 \\
HD 6582   & 5295 & 5413 & 5315 &  5408 \\
HD 6755   & 5053 & 5147 & 5008 &  5143 \\
HD 21581  & 4869 & 4844 & 4870 &  4932 \\
HD 25329  & 4777 & 4735 & 4842 &  4750 \\
HD 26297  & 4324 & 4420 & 4322 &  4390 \\
HD 29574  & 4047 & 4041 & 4020 &  4131 \\
HD 37828  & 4390 & 4448 & ---- &  4326 \\ 
HD 103095 & 5045 & 4924 & 5029 &  5000 \\
HD 165195 & 4237 & 4229 & 4237 &  4293 \\
HD 187111 & 4249 & 4200 & 4271 &  4433 \\
HD 206739 & 4653 & 4609 & 4647 &  4680 \\
HD 216143 & 4494 & 4433 & 4486 &  4360 \\
HD 221170 & 4412 & 4348 & 4410 &  4460 \\
BD +060648 & 4261 & 4217 & 4186 &  4308 \\
\enddata
\label{teffs}
\end{deluxetable}

\begin{deluxetable}{llllllllllllll}
\tablenum{4}
\tablewidth{0pt}
\footnotesize
\tablecaption{Adopted atmospheric parameters obtained from (a) spectroscopic
analysis (columns 2-5) and (b) IRFM temperatures,
trigonometric gravities and [\ion{Fe}{2}/H] (columns 6-9), compared 
with (c)  literature values (columns 10$-$14)}
\tablehead{
\multispan1 Star & \multispan4 {(a) Spectroscopic parameters} & 
\multispan4 {(b) IRFM, trigonometric log $g$ and \ion{Fe}{2}} & 
\multispan5 {(c) Literature parameters} \\
\colhead{}  &
\colhead{\teff$^{\rm sp}$} &
\colhead{log $g^{\rm sp}$}  &
\colhead{[Fe/H]$^{\rm sp}$} &
\colhead{v$_t$} &
\colhead{\teff$^{\rm IRFM}$} &
\colhead{log $g^{\rm Hip}$}  &
\colhead{[\ion{Fe}{2}/H]} &
\colhead{v$_t$} &
\colhead{\teff} &
\colhead{log $g$}  &
\colhead{[Fe/H]} &
\colhead{v$_t$} &
\colhead{ref.} 
}
\startdata
HD 2665    & 5030 & 2.20 & $-$1.90 & 1.4 & 4990 & 1.72$\pm$0.34 &$-$2.00& 1.4& 5000 & 2.20 & $-$1.97 & 1.3  & 1 \\
           &      &      &       &     & & & &  & 5050 & 2.20 & $-$1.96 & 1.6  & 2 \\
HD 6582    & 5408 & 4.40 & $-$0.83 & 0.8 & 5315 &4.52$\pm$0.00 &$-$0.80& 0.8& 5250 & 4.40 & $-$0.98 & 0.9  & 2 \\
           &      &      &       &     & & & &  & 5350 & 4.47 & $-$0.92 & 1.0  & 3 \\
HD 6755    & 5143 & 2.50 & $-$1.49 & 1.4 &5008 &2.93$\pm$0.10 &$-$1.48 & 1.4 & 5150 & 2.70 & $-$1.57 & 1.4  & 1 \\
           &      &      &       &     & & & &  & 5150 & 2.80 & $-$1.58 & 1.5  & 2 \\
HD 21581   & 4932 & 2.20 & $-$1.68 & 1.4 &4870 &2.71$\pm$0.25 &$-$1.45 & 1.4 & 4825 & 2.00 & $-$1.74 & 1.4  & 2 \\
           &      &      &       &     & & & &  & 4875 & 2.00 & $-$1.74 & 1.5  & 4 \\
HD 25329   & 4750 & 4.70 & $-$1.75 & 0.7 &4842 &4.83$\pm$0.02 &$-$1.72 & 0.7 & 4700 & 4.80 & $-$1.85 & 1.3  & 2 \\
           &      &      &       &     & & & &  & 4775 & 4.71 & $-$1.82 & 1.0  & 3 \\
           &      &      &       &     & & & &  & 4750 & 4.65 & $-$1.78 & 0.5  & 5 \\
           &      &      &       &     & & & &  & 4750& 4.70 & $-$1.74 & 1.1  & 6 \\              
HD 26297   & 4390 & 1.00 & $-$1.75 & 1.6 &4322 &0.94$\pm$0.93 &$-$1.71 & 1.6 & 4400 & 1.10 & $-$1.87 & 2.0  & 1 \\
           &      &      &       &     & & & &  & 4500 & 1.20 & $-$1.73 & 1.7  & 2 \\
           &      &      &       &     & & & &  & 4450 & 1.20 & $-$1.72 & 1.7  & 4 \\
HD 29574   & 4131 & 0.20 & $-$1.84 & 1.8 &4020 &0.30$\pm$0.82 &$-$1.76 & 1.8 & 4100 & 0.00 & $-$1.92 & 1.7  & 1 \\
           &      &      &       &     && & &  & 4300 & 0.40 & $-$1.88 & 1.7  & 2 \\
           &      &      &       &     & & & &  &4280 & 0.22 & $-$1.95 & 1.7  & 3 \\
           &      &      &       &     & & & &  &4100 & 0.00 & $-$1.93 & 1.6  & 7 \\
HD 37828   & 4326 & 1.05 & $-$1.42 & 1.3 & 4390 &1.58$\pm$0.19 &$-$1.25 & 1.3 &4350 & 1.20 & $-$1.42 & 1.4  & 7 \\
HD 103095  & 5000 & 4.30 & $-$1.46 & 1.1 &5029 &4.73$\pm$0.01 &$-$1.25 & 1.1 & 4950 & 4.50 & $-$1.46 & 0.7  & 2 \\
           &      &      &       &     & & & &  &5040 & 4.31 & $-$1.48 & 1.0  & 3 \\
           &      &      &       &     & & & &  & 5050 & 4.65 & $-$1.22 & 1.4  & 8 \\
HD 165195  & 4293 & 0.50 & $-$2.17 & 1.7 &4237 &1.05$\pm$0.45 &$-$2.01 & 1.7 &4450 & 1.10 & $-$2.24 & 1.9  & 1 \\
HD 187111  & 4433 & 1.14 & $-$1.56 & 1.6 &4271 &1.21$\pm$0.53 &$-$1.44 & 1.6 &4450 & 1.10 & $-$1.71 & 1.9  & 2 \\
           &      &      &       &     & & & &  &4400 & 1.08 & $-$1.72 & 1.7  & 3 \\
           &      &      &       &     & & & &  &4280 & 1.00 & $-$1.67 & 1.5  & 7 \\
           &      &      &       &     & & & &  &4250 & 0.70 & $-$1.74 & 1.7  & 1 \\   
HD 206739  & 4680 & 1.50 & $-$1.55 & 1.4 &4647 &0.97$\pm$0.89 &$-$1.70 & 1.4 &4675 & 1.70 & $-$1.58 & 1.7  & 1 \\
           &      &      &       &     & & & &  &4680 & 1.70 & $-$1.58 & 1.5  & 4 \\
HD 216143  & 4360 & 0.50 & $-$2.22 & 1.8 &4486 &1.80$\pm$0.35 &$-$1.75 & 1.8 &4525 & 0.80 & $-$2.18 & 1.8  & 1 \\
           &      &      &       &     & & & &  &4525 & 1.00 & $-$2.25 & 2.8  & 2 \\
           &      &      &       &     & & & &  &4400 & 0.70 & $-$2.26 & 1.8  & 7 \\
HD 221170  & 4460 & 0.75 & $-$2.11 & 1.6 &4410 &1.49$\pm$0.33 &$-$1.79 & 1.6 &4425 & 1.00 & $-$2.15 & 1.5  & 1 \\
           &      &      &       &     & & & &  &4500 & 0.90 & $-$2.19 & 2.7  & 2 \\
           &      &      &       &     & & & &  &4550 & 1.30 & $-$2.10 & 2.0  & 7 \\
BD +060648 & 4308 & 0.50 & $-$2.10 & 1.7 &4186 &1.87$\pm$0.46 &$-$1.24 & 1.7 &4430 & 1.00 & $-$2.11 & 1.8  & 7 \\
           &      &      &       &     & & & &  &4500 & 1.10 & $-$2.10 & 2.0  & 1 \\
\enddata
\label{tab4}
\tablerefs{(1) Pilachowski et al. 1996; (2) Fulbright 2000; (3) Tomkin \& Lambert 1999; 
(4) Kraft et al. 1992; (5) Beveridge \& Sneden 1994; (6) Spiesman \& Wallerstein 1991;
(7) Shetrone 1996; (8) Balachandran \& Carney 1996
 }
\end{deluxetable}

\begin{deluxetable}{llllllllllllllllllll}
\tablenum{5}
\tablewidth{0pt}
\tablecaption{Equivalent widths of IR OH and \ion{Fe}{1} lines}
\tablecomments{Identification of OH lines is given by
their (v',v"), branch, and rotational quantum number J".
* denotes blend of two \ion{Fe}{1} lines.
}
\tablerefs{
(1) HD 2665 (2) HD 6582 (3) HD 6755 (4) HD 21581 (5) HD 25329 (6) HD 26297
 (7) HD 29574 (8) HD 37828 (9) HD 103095  (10) HD 165195 (11) HD 187111
 (12) HD 206739 (13) HD 216143 (14) HD 221170 (15) BD +060648
 }
\footnotesize
\tablehead{
\colhead{OH,\ion{Fe}{1}}  &
\colhead{{$\lambda$}}   &
\colhead{{$\chi_{exc}$}}  &
\colhead{log {\it gf}} &
\colhead{1} &
\colhead{2} &
\colhead{3} &
\colhead{4} &  
\colhead{5} &  
\colhead{6} &  
\colhead{7} &  
\colhead{8} &  
\colhead{9} &  
\colhead{10} &  
\colhead{11} &  
\colhead{12} &  
\colhead{13} &  
\colhead{14} &  
\colhead{15} \\
\colhead{}  & 
\colhead{({\rm \AA})}   &
\colhead{(eV)}  &
\colhead{} &
\colhead{} &
\colhead{} &
\colhead{} &
\colhead{} &  
\colhead{} &  
\colhead{} &  
\colhead{} &  
\colhead{} &  
\colhead{} &  
\colhead{} & 
\colhead{} &  
\colhead{} &  
\colhead{} &  
\colhead{} &  
\colhead{} }
\startdata
3-1P1e5.5       & 15535.46 & 0.51 & $-$5.23 & -- & -- & -- & -- & 11 & 26 & 24 & 37 & -- & 24 & 38 & 10 & 11 & 13 & 12 \\ 
3-1P1f5.5       & 15536.71 & 0.51 & $-$5.23 & -- & -- & -- & -- & -- & 25 & 22 & -- & -- & -- & 37 & -- & -- & 14 & -- \\
2-0P2e10.5      & 15560.23 & 0.30 & $-$5.31 &  4 &  4 &  7 & 10 & 24 & 54 & 75 & 70 & 18 & 50 & 56 & -- & 24 & 20 & 36 \\
4-2R2e,f2.5     & 15565.9  & 0.90 & $-$5.00 & -- & -- & -- & -- & 11 & -- & 21 & 20 & -- & 13 & 20 & -- & -- & -- & 12 \\
2-0P1e11.5      & 15568.77 & 0.30 & $-$5.27 &  3 &  6 &  5 & 11 & 26 & 51 & 76 & 68 & 14 & 46 & 56 & 20 & 22 & 20 & 40 \\
2-0P1f11.5      & 15572.07 & 0.30 & $-$5.27 &  7 &  5 &  5 & 12 & 26 & 50 & 73 & 66 & 18 & 49 & 58 & 20 & 26 & -- & 39 \\
\ion{Fe}{1}     & 15534.26 & 5.64 & $-$0.47 & -- & 42 & -- & 15 & -- & 27 & 23 & 48 & -- &  6 & 35 & 26 &  7 & 14 & 10 \\
\ion{Fe}{1}$^*$ & 15550.45 & 6.34 & $-$0.35 & -- & 11 &  7 & -- & -- &  7 & -- & 17 & -- & -- & -- & -- & -- & -- & -- \\
\ion{Fe}{1}     & 15551.43 & 6.35 & $-$0.31 & -- & 13 &  7 & -- & -- &  7 & -- & 19 & -- & -- &  8 & -- & -- & -- & -- \\
\ion{Fe}{1}$^*$ & 15591.49 & 6.30 & $\phantom{-}$0.66 & 21 & 94 & 50 & -- & -- & -- & -- & -- & -- & 26 & 74 & 61 & 21 & 30 & 31 \\
\ion{Fe}{1}     & 15604.22 & 6.24 & $\phantom{-}$0.28 &  9 & -- & 20 & -- & -- & -- & -- & -- & -- & -- & 40 & -- & -- & -- & 20 \\
\enddata
\label{tab5}
\end{deluxetable}

\begin{deluxetable}{lll}
\tablenum{6}
\tablewidth{0pt}
\footnotesize
\tablecaption{[C/Fe] from CO lines (spectroscopic parameters) and literature values}
\tablecomments{Solar abundances from Grevesse et al. (1996): 
$\epsilon$(Fe) = 7.50 and $\epsilon$(C) = 8.55 (in the scale log(X/H) + 12)}
\tablehead{
\colhead{Star}  &
\colhead{[C/Fe]$_{\rm CO}$} &
\colhead{[C/Fe]$_{\rm lit}$} }
\startdata
HD 2665   &         & $-$0.30$^{1, 2}$ \\
HD 6582   & +0.0    & +0.04$^3$ \\
HD 6755   &         & $-$0.10$^{1, 2}$ \\
HD 21581  &         & $-$0.40$^2$ \\
HD 25329  &         & +0.10$^4$ \\
HD 26297  & $-$0.2  &         \\
HD 29574  & $-$1.0  & $<$$-$1.0$^2$ \\
HD 37828  & $-$0.35 &      \\
HD 103095 & $-$0.5  & $-$0.32$^5$,$-$0.5$^6$ \\
HD 165195 &         & $-$1.00$^{1,6}$  \\
HD 187111 & $-$0.6  &       \\
HD 206739 & $-$0.5  & $-$0.50$^1$   \\
HD 216143 &         & $-$0.65$^1$  \\
HD 221170 &         & $-$0.30$^{1,6}$  \\
BD +060648 &        & $-$0.70$^{1, 2}$ \\
\enddata
\label{tab6}
\tablerefs{(1) Kraft et al. 1982; (2) Friel et al. 1982; (3) Clegg, Lambert 
\& Tomkin 1981; (4) Tomkin et al. 1992; (5) BC96; 
(6) Sneden, Pilachowski \& VandenBerg 1986}
\end{deluxetable}

\begin{deluxetable}{llllllll}
\tablenum{7a}
\tablewidth{0pt}
\footnotesize
\tablecaption{Metallicities obtained from optical \ion{Fe}{1}, \ion{Fe}{2} lines
and from infrared \ion{Fe}{1} lines, using (a) Spectroscopic parameters,
(b) IRFM \teff and trigonometric gravities.
}
\tablehead{
\multispan1 Star & \multispan2 {(a) \teff$^{\rm spec}$, log $g^{\rm Hip}$} &
\multispan2 {(b) IRFM \teff, log $g^{\rm Hip}$} \\
\colhead{}  &
\colhead{[Fe/H]$_{\rm opt}$} &
\colhead{[Fe/H]$_{\rm IR}$} &
\colhead{\rm [Fe II/H]$_{\rm opt}$} &
\colhead{[Fe/H]$_{\rm IR}$} & }
\startdata
HD 2665   &  $-$1.90 & $-$1.88 & $-$2.00 & $-$1.91  \\
HD 6582   &  $-$0.83 & $-$0.93 & $-$0.80 & $-$0.95  \\
HD 6755   &  $-$1.49 & $-$1.39 & $-$1.48 & $-$1.42  \\
HD 21581  &  $-$1.68 & $-$1.66 & $-$1.45 & $-$1.69  \\
HD 25329  &  $-$1.75 &  --     & $-$1.72 &	--  \\
HD 26297  &  $-$1.75 & $-$1.64 & $-$1.71 & $-$1.67  \\
HD 29574  &  $-$1.84 & $-$1.83 & $-$1.76 & $-$1.92  \\
HD 37828  &  $-$1.42 & $-$1.27 & $-$1.25 & $-$1.23  \\
HD 103095 &  $-$1.46 &    --   & $-$1.25 &	--  \\
HD 165195 &  $-$2.17 & $-$2.19 & $-$2.01 & $-$2.29  \\
HD 187111 &  $-$1.56 & $-$1.51 & $-$1.44 & $-$1.59  \\
HD 206739 &  $-$1.55 & $-$1.47 & $-$1.70 & $-$1.48  \\
HD 216143 &  $-$2.22 & $-$2.17 & $-$1.75 & $-$2.17  \\
HD 221170 &  $-$2.11 & $-$1.92 & $-$1.79 & $-$1.97  \\
BD +060648 & $-$2.10 & $-$1.96 & $-$1.24 & $-$2.03  \\
\enddata
\label{}
\end{deluxetable}

\begin{deluxetable}{lllllllllll}
\tablenum{7b}
\tablewidth{0pt}
\footnotesize
\tablecaption{Oxygen abundances from IR OH lines: (a) spectroscopic parameters,
(b) IRFM \teff, trigonometric gravities, \ion{Fe}{2} and (c) mean values
}
\tablehead{
\multispan1 Star & \multispan3 {(a) spectroscopic parameters} & 
\multispan3 {(b) IRFM \teff, log $g^{\rm Hip}$, \ion{Fe}{2}} &
\multispan3 {(c) mean values}\\
\colhead{}  &
\colhead{[Fe/H]$^{\rm sp}$} &
\colhead{[O/H]$^{\rm sp}$} &
\colhead{[O/Fe]$^{\rm sp}$} &
\colhead{[\ion{Fe}{2}/H]} &
\colhead{[O/H]$^{\rm Hip}$} &
\colhead{[O/Fe]$^{\rm Hip}$} &
\colhead{[$\overline{\rm Fe/H}$]} &
\colhead{[$\overline{\rm O/H}$]} &
\colhead{[$\overline{\rm O/Fe}$]} }
\startdata
HD 2665   & $-$1.90 & $-$1.35 & 0.55 & $-$2.00 & $-$1.31 & 0.69 & $-$1.95 & $-$1.33 & 0.62 \\
HD 6582   & $-$0.83 & $-$0.58 & 0.25 & $-$0.80 & $-$0.57 & 0.23 & $-$0.82 & $-$0.58 & 0.24 \\
HD 6755   & $-$1.49 & $-$1.09 & 0.40 & $-$1.48 & $-$1.28 & 0.20 & $-$1.49 & $-$1.19 & 0.30 \\
HD 21581  & $-$1.68 & $-$1.20 & 0.48 & $-$1.45 & $-$1.30 & 0.15 & $-$1.57 & $-$1.25 & 0.32 \\
HD 25329  & $-$1.75 & $-$1.30 & 0.45 & $-$1.72 & $-$1.21 & 0.51 & $-$1.74 & $-$1.26 & 0.48 \\
HD 26297  & $-$1.75 & $-$1.31 & 0.44 & $-$1.71 & $-$1.41 & 0.30 & $-$1.73 & $-$1.36 & 0.37 \\
HD 29574  & $-$1.84 & $-$1.29 & 0.55 & $-$1.76 & $-$1.51 & 0.25 & $-$1.80 & $-$1.40 & 0.40 \\
HD 37828  & $-$1.42 & $-$0.88 & 0.54 & $-$1.25 & $-$0.85 & 0.40 & $-$1.34 & $-$0.87 & 0.47 \\
HD 103095 & $-$1.46 & $-$1.09 & 0.37 & $-$1.25 & $-$1.02 & 0.23 & $-$1.36 & $-$1.06 & 0.30 \\
HD 165195 & $-$2.17 & $-$1.49 & 0.68 & $-$2.01 & $-$1.77 & 0.24 & $-$2.09 & $-$1.63 & 0.46 \\
HD 187111 & $-$1.56 & $-$0.91 & 0.65 & $-$1.44 & $-$1.22 & 0.22 & $-$1.50 & $-$1.06 & 0.44 \\
HD 206739 & $-$1.55 & $-$1.08 & 0.47 & $-$1.70 & $-$1.13 & 0.57 & $-$1.63 & $-$1.11 & 0.52 \\
HD 216143 & $-$2.22 & $-$1.61 & 0.61 & $-$1.75 & $-$1.69 & 0.06 & $-$1.99 & $-$1.65 & 0.34 \\
HD 221170 & $-$2.11 & $-$1.63 & 0.48 & $-$1.79 & $-$1.79 & 0.00 & $-$1.95 & $-$1.71 & 0.24 \\
BD +060648 & $-$2.10 & $-$1.48 & 0.62 & $-$1.24 & $-$1.62 & $-$0.38& $-$1.67 & $-$1.55 & 0.12 \\
\enddata
\label{}
\end{deluxetable}

\begin{deluxetable}{cccccc}
\tablenum{8}
\tablewidth{0pt}
\footnotesize
\tablecaption{Sensitivity to  \teff, log {\it g} and v$_t$  for 
giants (a mean computed for HD 37828 and HD 216143) and
dwarfs (computed for HD 103095).}
\tablehead{
\colhead{Abundance}  &
\colhead{$\Delta$ \teff} &
\colhead{$\Delta$ log g} &
\colhead{$\Delta$ v$_t$} &
\colhead{($\Sigma$x$^2$)$^{1/2}$} \\
\colhead{}  &
\colhead{+100 K} &
\colhead{+0.5 dex} &
\colhead{+0.5 km s$^{-1}$} &
\colhead{}
}
\startdata
\multispan5 Giants \\
{[FeI/H]}  & +0.13 & +0.05   & $-$0.15 & 0.20 \\
{[FeII/H]} & $-$0.10 & +0.22 & $-$0.04 & 0.24 \\
{[O/H]}    & +0.20 & $-$0.10 & +0.03  & 0.23 \\
{[O/FeI]}  & +0.07 & $-$0.15 & +0.18  & 0.24 \\
{[O/FeII]} & +0.30 & $-$0.32 & +0.07  & 0.44 \\
\multispan5 Dwarfs \\
{[FeI/H]}  & +0.07 & +0.01   & $-$0.05 & 0.09 \\
{[FeII/H]} & $-$0.03 & +0.18 & $-$0.01 & 0.18 \\
{[O/H]}    & +0.17 & $-$0.03 & +0.03  & 0.18 \\
{[O/FeI]}  & +0.10 & +0.02   & +0.08  & 0.13 \\
{[O/FeII]} & +0.20 & $-$0.15 & +0.04  & 0.25 \\
\enddata
\label{tabxxxxx}
\end{deluxetable}

\begin{deluxetable}{llllll}
\tablenum{9}
\tablewidth{0pt}
\footnotesize
\tablecaption{[O/Fe] values from the literature compared to present results
(mean values are given, see Sect. 3)}
\tablehead{
\colhead{Star}  &
\colhead{[\ion{O}{1}]} &
\colhead{\ion{O}{1}}  &
\colhead{UV OH} &
\colhead{IR OH} &
\colhead{reference} }
\startdata
HD 2665    &      &      &       & 0.62 & 1  \\
           & 0.20 & 0.28 &       &      & 2  \\
HD 6582    &      &      &       & 0.24 & 1  \\
           &      & 0.45 &       &      & 3   \\ 
           &      &      & 0.28  &      & 4  \\
HD 6755    &      &      &       & 0.30 & 1   \\
HD 21581   &      &      &       & 0.32 & 1  \\
           & 0.29 &      &       &      & 5   \\
           & 0.25 &      &       &      & 6  \\
           & 0.39 & 0.32 &       &      & 2  \\
HD 25329   &      &      &       & 0.48 & 1  \\
           & 0.31 &      &       &      & 7  \\
           &      & 1.07 &       &      & 8  \\
HD 26297   &      &      &       & 0.37 & 1  \\
           & 0.39 &      &       &      & 5   \\
           & 0.25 &      &       &      & 6  \\
           & 0.45 &      &       &      & 9  \\
           & 0.31 & 0.41 &       &      & 2  \\
HD 29574   &      &      &       & 0.40 & 1  \\
           & 0.30 &      &       &      &  10   \\
           & 0.50 &      &       &      & 6  \\
HD 37828   &      &      &       & 0.47 & 1  \\
           & 0.35 &      &       &      & 10   \\
           & 0.32 &      &       &      & 11  \\
HD 103095  &      &      &       & 0.30 & 1  \\
           & 0.48 &      &       &      & 12 \\
           &      & 0.89 &       &      & 13   \\
           & 0.33 & 0.42 &       & 0.29 & 14   \\
           & 0.37 &      & 0.62  &      & 4  \\
           &      & 0.40 & 0.44  &      & 15 King scale \\
           &      & 0.61 & 0.29  &      & 15 Carney scale \\
           & 0.45 & 0.22 &       &      & 2  \\
HD 165195  &      &      &       & 0.46 & 1  \\
           & 0.59 & 0.61 &       &      & 16  \\
           & 0.40 &      &       &      & 6  \\
HD 187111  &      &      &       & 0.44 & 1  \\
           & 0.34 &      &       &      & 5  \\
           & 0.40 &      &       &      & 6  \\   
           & 0.34 &      &       &      & 7  \\   
           & 0.35 &      &       &      & 2  \\
HD 206739  &      &      &       & 0.52 & 1  \\
           & 0.25 &      &       &      & 5   \\
           & 0.25 &      &       &      & 6  \\
HD 216143  &      &      &       & 0.34 & 1  \\
           & 0.16 &      &       &      & 10   \\
           & 0.69 &      &       &      & 9 \\
HD 221170  &      &      &       & 0.24 & 1  \\
           & 0.22 &      &       &      & 5   \\
BD +060648 &      &      &       & 0.12 & 1  \\
           & 0.45 &      &       &      & 10  \\
\enddata
\label{tabxxxx}
\tablerefs{(1)  This work; (2) Gratton et al. (2000); 
(3) Clegg et al. 1981; (4) IGR98; (5) Kraft et al. 1992;  
(6) Barbuy (1988); (7) Spiesman \& Wallerstein 1991;
(8) Beveridge \& Sneden 1994; (9) Gratton \& Ortolani 1986;
(10) Shetrone (1996); (11) Barbuy \& Erdelyi-Mendes (1989);
(12) Spite \& Spite (1991); (13) Tomkin et al. 1992; 
(14) BC96; (15) BKDV99; 
(16) Takeda et al. (2000).
 }
\end{deluxetable}

\begin{deluxetable}{ccccc}
\tablenum{10}
\tablewidth{0pt}
\footnotesize
\tablecaption{[O/Fe] ratios for HD 103095 using different model parameters}
\tablecomments{Oxygen abundances based on IR OH lines, [\ion{O}{1}] and 
\ion{O}{1} lines were derived 
using data by BC96, Spite \& Spite (1991) and Tomkin et al. (1992), respectively. 
For the UV OH lines  the [O/Fe] ratios are reported as given in 
IGR98 and BKDV99}
\tablehead{
\colhead{Model atmosphere}  &
\colhead{IR OH} &
\colhead{UV OH} &
\colhead{[\ion{O}{1}]} &
\colhead{\ion{O}{1}} \\
\colhead{\teff, log {\it g}, [Fe/H], v$_t$, reference}  &
\colhead{} &
\colhead{} &
\colhead{} &
\colhead{}
}
\startdata
5000 K, 4.30, $-$1.46, 1.1 km s$^{-1}$, 1a & +0.37 &   ---    & +0.31 & +0.60 \\
5029 K, 4.73, $-$1.25, 1.1 km s$^{-1}$, 1b & +0.23 &   ---    & +0.34 & +0.50 \\
5040 K, 4.65, $-$1.22, 1.4 km s$^{-1}$, 2 & +0.28 &   ---     & +0.28 & +0.42 \\
5030 K, 4.55, $-$1.30, 1.0 km s$^{-1}$, 3 & +0.28 & +0.62$^3$ & +0.29 & +0.48 \\
5005 K, 4.65, $-$1.37, 1.5 km s$^{-1}$, 4 & +0.30 & +0.29$^4$ & +0.37 & +0.62 \\
5132 K, 4.75, $-$1.26, 1.5 km s$^{-1}$, 5 & +0.39 & +0.44$^5$ & +0.37 & +0.40 \\
\enddata
\label{tab7}
\tablerefs{(1) This work; (2) BC96; 
(3) IGR98; (4) BKDV99 (Carney scale); (5) BKDV99 (King scale) }
\end{deluxetable}